\documentclass[showpacs, preprintnumbers,superscriptaddressprl,floatfix,letterpaper,groupedaddress]{revtex4}
\usepackage{graphicx}
\usepackage{amsfonts}
\usepackage{amsmath, amsthm, amssymb}

\def \pslash {p\!\!\!/}				% definition of slash character
\begin{document}
\title[Neutrino matter potentials induced by Earth]{Neutrino matter potentials induced by Earth}
\author{J. Linder}
\email{jacob.linder@ntnu.no}
\affiliation{Department of Theoretical Physics, Norwegian University of
Science and Technology, N-7491 Trondheim, Norway}
\date{Received \today}
\begin{abstract}
An instructive method of deriving the matter potentials felt by neutrinos propagating through matter on Earth is presented. This paper thoroughly guides the reader through the calculations involving the effective weak Hamiltonian for lepton and quark scattering. The matter potentials are well-known results since the late 70's, but a detailed and pedagogical calculation of these quantities is hard to find. We derive potentials due to charged and neutral current scattering on electrons, neutrons and protons. Intended readership is for undergraduates/graduates in the fields of relativistic quantum mechanics and quantum field theory. In addition to the derivation of the potentials for neutrinos, we explicitely study the origin of the reversed sign for potentials in the case of antineutrino-scattering. 
\end{abstract}
\maketitle
\section{Introduction}
In his famous paper from 1978 \cite{wolf}, Wolfenstein showed that neutrino oscillations \cite{msw2} are altered in the presence of matter; a phenomenon which came to be known as the MSW-effect after Mikheyev, Smirnov, and Wolfenstein. This occurs as a result of neutrinos experiencing potentials due to charged and neutral current scattering on electrons, neutrons, and protons as they propagate through matter on Earth. The explicit expressions for these matter potentials can be found in much literature, and a brief sketch for how these quantities are derived is given in \cite{msw3}, which deals with the special case of charged current electron-neutrino scattering on electrons. Following the outlines of this sketch, this paper provides a thorough and pedagogical walk-through of calculating \mbox{\it all} relevant matter potentials that neutrinos experience when propagating through matter in Earth. The calculations include many techniques that are relevant when setting out to find expectation values in the second quantized formalism, and should hence be of interest for students in the process of being equipped to deal with scattering reactions in the framework of quantum field theory. The physics underlying the calculation is also explained to make the derivation more comprehensible, with particular emphasis on how the procedure of finding the potentials can be considerably simplified by observing that they are independent of the axial coupling constant, in addition to an analysis of why the potential sign is reversed when considering antineutrinos. Especially the latter argument is of particular value for the intended audience, since practically all literature simply states that the sign is reversed instead of \mbox{\it explaining} why it is reversed. 

\section{Neutrino eigenstates and effective Hamiltonian}
The vacuum Hamiltonian describing a propagating neutrino flavor state is as previously mentioned altered in the presence of matter. This is readily seen by considering the energy contribution to the Hamiltonian from scattering on nuclear components such as electrons and neutrons. First of all, the flavor eigenstates produced in weak interactions are coupled to the mass eigenstates through $[\nu_e\mbox{ }\nu_\mu\mbox{ }\nu_\tau]^{\mbox{\tiny T}} = U[\nu_1\mbox{ }\nu_2\mbox{ }\nu_3]^{\mbox{\tiny T}}$, where $U$ is the unitary mixing matrix usually parametrized as in Ref. \cite{msw2},

\begin{equation}\label{eq:Upara}
U = U_{23}U_{13}U_{12} = \left[ \begin{array}{ccc}
	1& 0 & 0 \\
	0 & c_{23} & s_{23} \\
	0 & -s_{23} & c_{23} \\
      \end{array} \right]\left[ \begin{array}{ccc}
	c_{13} & 0 & s_{13}\mbox{e}^{-\mbox{\scriptsize i}\delta_{\mbox{\tiny CP}}} \\
	0 & 1 & 0 \\
	-s_{13}\mbox{e}^{\mbox{\scriptsize i}\delta_{\mbox{\tiny CP}}} & 0 & c_{13} \\
      \end{array} \right]\left[ \begin{array}{ccc}
	c_{12} & s_{12} & 0\\
	-s_{12} & c_{12} & 0 \\
	0& 0 &1\\
      \end{array} \right].
\end{equation}

Here, $s_{ij}=\sin\theta_{ij}$ and $c_{ij}=\cos\theta_{ij}$ and $\theta_{ij}$ are the mixing angles between the neutrino 
states.  The $\delta_{\mbox{\tiny CP}}$ is the $CP$-violating phase with allowed values $\delta_{\mbox{\tiny CP}}\in [0,2\pi]$. 
For $\delta_{\mbox{\tiny CP}} \in \{0,\pi\}$ we have $CP$ invariance, while the violation is at its largest when 
$\delta_{\mbox{\tiny CP}} \in \{\pi/2,3\pi/2\}$. For our purposes, $\delta_{\mbox{\tiny CP}}$ is set to zero. This is motivated by the fact that
$CP$-violation in the standard model requires a non-zero value of $\theta_{13}$, and since experiments have placed stringent limits on the value of $\theta_{13}$, the effective $CP$-violation is quite small (see {\it e.g.} Ref. \cite{msw3}). Note that we have intentionally left out diagonal Majorana phases that are of no consequence for oscillation experiments.\\

The evolution of the mass eigenstates $\Psi^{\mbox{\tiny M}}(x)$ in vacuum is described by 
\begin{equation}
\mbox{i}\frac{\mbox{d}}{\mbox{d}t}\Psi^{\mbox{\tiny M}}(x) = H_0^{\mbox{\tiny M}}\Psi^{\mbox{\tiny M}}(x),
\end{equation}
where $H_0 = \mbox{diag}(E_1,E_2,E_3)$. $E_i, i=1,2,3$, are the energy eigenvalues for mass eigenstates. To obtain the Hamiltonian in flavor space, we perform $H_{0} = UH_0^{\mbox{\tiny M}}U^{-1}$, where $U$ is the mixing matrix from Eq. (\ref{eq:Upara}). When the neutrinos propagate through matter, we must add the contributions from scattering on particles present. Accordingly, one obtains $H = H_0 + H_{\mbox{\scriptsize I}}$, written out as
\begin{align}\label{eq:hmsw}
H_{\mbox{\scriptsize I}} &= H_Z^n + H_Z^p + H_Z^e + H_W^e,
\end{align}
where the interaction Hamiltonians are given as $H_Z^i = \mbox{diag}(V_Z^i,V_Z^i,V_Z^i)$, $i=n,p,e$, and $H_W^e = \mbox{diag}(V_W^e,0,0)$. The superscript refers to the scattering component while the subscript indicates which gauge boson mediates the reaction, i.e. neutral or charged current. For example, $V_Z^e$ represents the effective matter potential due to NC (neutral-current) scattering on electrons. Note that we have only included matter potentials arising from reactions with electrons, protons, and neutrons, since the the concentrations of $\mu$ and $\tau$ particles is virtually zero on Earth. \\

In deriving the matter potentials, we are going to need the effective weak CC (charged-current) and NC interaction Hamiltonians 
\begin{align}\label{hw}
{\cal{H}}^{\mbox{\scriptsize eff}}_W &= \frac{G_F}{\sqrt{2}} J_{W\mu} J^{\mu\dag}_W,
\end{align}
\begin{align}\label{hz}
{\cal{H}}^{\mbox{\scriptsize eff}}_Z &= \frac{4G_F}{\sqrt{2}} J^\mu_Z J_{Z\mu}.
\end{align}
The Fermi-constant is $G_F = \sqrt{2}g^2/8m_W^2$, while the currents are defined as 
\begin{align}\label{modcurrents}
J^\mu_W &=  \sum_l \bigl[\overline{l} \gamma^\mu (1-\gamma_5) \nu_l + \overline{d}_{\theta} \gamma^\mu (1-\gamma_5)u \bigl], \notag \\ 
J^\mu_Z &= \frac{1}{2} \sum_i \overline{\psi}_i \gamma^\mu \Bigl[I^3_i(1-\gamma_5) - 2Q_i\sin^2\theta_W\Bigr]\psi_i,
\end{align}
where $i = (l,\nu_l,u,d)$, while $I^3_i$ is the accompanying particle isospin and $Q_i$ is the particle charge in units of $e$. Now, $d_{\theta}$=$d\cos{\theta_C} + s\sin{\theta_C}$. The Cabbibo angle $\theta_C$ has been experimentally determined to $\cos\theta_C \approx 0.98$, so we shall exclude the $s$ quark part from now on and set $d_{\theta} = d$. This notation has been adapted from Refs. \cite{qft, quigg}. \\

The derivation of neutrino matter potentials is certainly not new, and has been studied in {\it e.g.} Refs. \cite{wolf,langacker,ms,bethe}. This paper contributes to the study of matter potentials by presenting an instructive method to derive the matter potentials that applies in the same way to leptons as nucleons. 

\section{Derivation of $V_W^e$}\label{VWe}
First, we seek the quantity 
\begin{equation}\label{vcc}
V_W^e = \langle \nu_e(p_1,s_1)e(p_2,s_2)|H_W|\nu_e(p_1,s_1)e(p_2,s_2) \rangle,
\end{equation}
where $H_W$ is the charged-current contribution due to scattering on electrons, omitting the superscript $^{\mbox{\scriptsize eff}}$ for clarity. Since we are dealing with elastic scattering, it is fair to assume that the neutrinos and electrons conserve their momentum as shown in Eq. (\ref{vcc}). The low-energy effective Hamiltonian density relevant for CC $\nu_ee$ scattering is found in Eq. (\ref{hw}), namely 
\begin{align}\label{hcc0}
{\cal{H}}_W(x) &= \frac{G_F}{\sqrt{2}}\bigl[\overline{e}(x)\gamma^\beta(1-\gamma_5)\nu_e(x)\bigr]\bigl[\overline{\nu}_e(x)\gamma_\beta(1-\gamma_5)e(x)\bigr].
\end{align}
Naïve insertion of this density into
\begin{equation}
H_W = \int_V {\cal{H}}_W(x) \mbox{d}{\bf x},
\end{equation}
is not correct. The presence of electrons in a medium leads to two important modifications. First of all, the statistical energy distribution of the electrons in the medium is accounted for by integration over the Fermi function $f(E_e,T)$ which is normalized to $\int f(E_e,T) \mbox{d}{\bf p}_e = 1$. Secondly, since we do not know the polarization of the electrons, an averaging over spins $1/2 \sum_s$ is needed. In total, this corresponds to transforming Eq. (\ref{hcc0}) to
\begin{align}\label{hwe}
{\cal{H}}_W(x) &= \int  f(E_e,T) \frac{G_F}{\sqrt{2}} \times \frac{1}{2}  \sum_{s_2} \bigl[\overline{e}(x)\gamma^\beta(1-\gamma_5)\nu_e(x)\bigr]\notag\\
&\hspace{0.1in}\times\bigl[\overline{\nu}_e(x)\gamma_\beta(1-\gamma_5)e(x)\bigr] \mbox{d}{\bf p}_2.
\end{align}
Since only electrons with $(p,s) = (p_2,s_2)$ will contribute to Eq. (\ref{vcc}), we obtain

\begin{align}\label{vcc1}
\bigl[\overline{e}(x)\gamma^\beta(1-\gamma_5)\nu_e(x)\bigr]\bigl[\overline{\nu}_e&(x)\gamma_\beta(1-\gamma_5)e(x)\bigr]|\nu_e(p_1,s_1)e(p_2,s_2) \rangle  \notag  \\
= \frac{1}{2VE_2(p_2)}&\bigl[a_{s_2}^\dagger({\bf p}_2) a_{s_2}({\bf p}_2)\overline{u}_{s_2}(p_2)\gamma^\beta(1-\gamma_5)\nu_e(x)\bigr]\bigl[\overline{\nu}_e(x)\gamma_\beta(1-\gamma_5)u_{s_2}(p_2)\bigr]|\nu_e(p_1,s_1)e(p_2,s_2) \rangle.
\end{align}
Making the identification of the number operator $\mbox{Num}_{s_2}({\bf p}_2) = a_{s_2}^\dagger({\bf p}_2)a_{s_2}({\bf p}_2)$, insertion of Eq. (\ref{vcc1}) into Eq. (\ref{vcc}) produces
\begin{align}\label{vcc2}
V_W^e &= \langle \nu_e(p_1,s_1)e(p_2,s_2)| \frac{G_F}{4\sqrt{2}V}\times \int \int f(E_e,T) \sum_{s_2} \frac{\mbox{Num}_{s_2}({\bf p}_2)}{E_e({\bf p}_2)} \bigl[\overline{u}_{s_2}(p_2)\gamma^\beta(1-\gamma_5)\nu_e(x)\bigr]   \notag \\
&\times \bigl[\overline{\nu}_e(x)\gamma_\beta(1-\gamma_5)u_{s_2}(p_2)\bigr] \mbox{d}{\bf x} \mbox{ d}{\bf p}_2 |\nu_e(p_1,s_1)e(p_2,s_2) \rangle.
\end{align}
It should be clear that the $\nu_e(x)$ and $\overline{\nu}_e(x)$ symbols refer to the second quantized fields 
\begin{align}\label{lepfield}
\psi(x) = \sum_{s,\mathbf{p}} \sqrt{\frac{1}{2V\omega_{\mathbf{p}}}} \Big[ a_{s}(\mathbf{p})u_s(\mathbf{p})\mbox{e}^{-\mbox{\scriptsize i}px} + b_{s}^\dag(\mathbf{p})v_s(\mathbf{p})\mbox{e}^{\mbox{\scriptsize i}px}\Big] \notag \\
\overline{\psi}(x) = \sum_{s,\mathbf{p}} \sqrt{\frac{1}{2V\omega_{\mathbf{p}}}} \Big[ b_{s}(\mathbf{p})\overline{v}_s(\mathbf{p})\mbox{e}^{-\mbox{\scriptsize i}px} + a_{s}^\dag(\mathbf{p})\overline{u}_s(\mathbf{p})\mbox{e}^{\mbox{\scriptsize i}px}\Big],
\end{align}
where $\psi$ is a lepton or a quark. On the other hand, the state vectors $|\nu_e(p_i,s_i)\rangle$, $i=1,2$, are simply representations for neutrinos with 4-momentum $p_i$ and spin $s_i$. In order to continue without too many complications, assume that the material is isotropic and holds an equal number of electrons with spin up as spin down, i.e. a non-magnetic material. This leads to
\begin{align}\label{vcc3}
V_W^e &= \langle \nu_e(p_1,s_1)e(p_2,s_2)| \frac{G_F}{4\sqrt{2}}\times \int \int \frac{f(E_e,T) N_e({\bf p}_2)}{E_e({\bf p}_2)} \sum_{s_2}  \bigl[\overline{u}_{s_2}(p_2)\gamma^\beta(1-\gamma_5)\nu_e(x)\bigr] \notag \\
&\hspace{0.2in} \times \bigl[\overline{\nu}_e(x)\gamma_\beta(1-\gamma_5)u_{s_2}(p_2)\bigr] \mbox{d}{\bf x} \mbox{ d}{\bf p}_2 |\nu_e(p_1,s_1)e(p_2,s_2) \rangle \notag \\
&= \langle \nu_e(p_1,s_1)e(p_2,s_2)| \frac{G_F}{4\sqrt{2}}\times \int \int \frac{f(E_e,T) N_e({\bf p}_2)}{E_e({\bf p}_2)} \overline{\nu}_e(x)\gamma_\beta(1-\gamma_5)\nu_e(x) \notag \\
&\hspace{0.2in} \times  \sum_{s_2}  \bigl[\overline{u}_{s_2}(p_2)\gamma^\beta(1-\gamma_5))u_{s_2}(p_2)\bigr] \mbox{d}{\bf x} \mbox{ d}{\bf p}_2 |\nu_e(p_1,s_1)e(p_2,s_2) \rangle. 
\end{align}
Here, we have used the Fierz identity to re-arrange the $\nu_e$ and $e$ spinors in such a fashion that makes it possible to extract the neutrino-spinor part from the summation over $s_2$. The remaining sum is evaluated by
\begin{align}\label{trvcc}
\sum_{s_2}  \bigl[\overline{u}_{s_2}(p_2)\gamma^\beta(1-\gamma_5)u_{s_2}(p_2)\bigr] &= \mbox{Tr}\{(\pslash_2 + m_e)\gamma^\beta(1-\gamma_5) \} \notag \\
&= p_{2\alpha} \mbox{Tr}\{ \gamma^\alpha\gamma^\beta(1-\gamma_5) \} = 4p_2^\beta,
\end{align}
with efficient use of standard trace relations. Eq. (\ref{trvcc}) in synthesis with (\ref{vcc3}) leads to 
\begin{align}\label{vcc4}
V_W^e &= \langle \nu_e(p_1,s_1)e(p_2,s_2)| \frac{G_F}{\sqrt{2}}\times \int \int \frac{f(E_e,T) N_e({\bf p}_2)}{E_e({\bf p}_2)} \notag \\
&\hspace{0.2in} \times \overline{\nu}_e(x)\gamma_\beta(1-\gamma_5)\nu_e(x) p_2^\beta \mbox{d}{\bf x} \mbox{ d}{\bf p}_2 |\nu_e(p_1,s_1)e(p_2,s_2) \rangle \notag \\
&= \langle \nu_e(p_1,s_1)e(p_2,s_2)| \frac{G_FN_e}{\sqrt{2}}\times \int  \overline{\nu}_e(x)\gamma_0(1-\gamma_5)\nu_e(x)  \mbox{d}{\bf x}  |\nu_e(p_1,s_1)e(p_2,s_2) \rangle,
\end{align}
where we have exploited the isotropy $\int {\bf p}_2f(E_e,T)\mbox{ d}{\bf p}_2 = 0$ and the expression for the total electron density $\int f(E_e,T) N_e({\bf p}_2) \mbox{ d}{\bf p}_2 = N_e$. Only integration over {\bf x} remains, such that we obtain
\begin{align}\label{vcc5}
V_W^e &= \langle \nu_e(p_1,s_1)e(p_2,s_2)| \frac{G_FN_e}{\sqrt{2}}\times \int \overline{\nu}_e(x)\gamma_0(1-\gamma_5)\nu_e(x) \mbox{d}{\bf x}  |\nu_e(p_1,s_1)e(p_2,s_2) \rangle \notag \\
&= \langle \nu_e(p_1,s_1)e(p_2,s_2)|\frac{G_FN_e}{\sqrt{2}}\times \frac{1}{2VE_{\nu_e}} \int \mbox{Tr}\{ (\pslash_{\nu_e} + m_{\nu_e})\gamma^0(1-\gamma^5) \} \notag\\
&\hspace{0.2in} \times \overbrace{a_{s_1}^\dagger({\bf p}_1) a_{s_1}({\bf p}_1)}^{\mbox{\scriptsize Num}_{s_1}({\bf p}_1)}    \mbox{d}{\bf x} |\nu_e(p_1,s_1)e(p_2,s_2) \rangle\notag \\
&= \langle \nu_e(p_1,s_1)e(p_2,s_2)|\frac{G_FN_e}{\sqrt{2}}\times \frac{1}{2VE_{\nu_e}} \int 4E_{\nu_e} \mbox{d}{\bf x} |\nu_e(p_1,s_1)e(p_2,s_2)\rangle.
\end{align}
Assuming normalized state vectors $|\nu_e(p_1,s_1)e(p_2,s_2)\rangle$, Eq. (\ref{vcc5}) reduces to 
\begin{equation}
V_W^e = \frac{G_FN_e}{\sqrt{2}}\times \frac{2}{V} \int \mbox{d}{\bf x} \langle \nu_e(p_1,s_1)e(p_2,s_2)||\nu_e(p_1,s_1)e(p_2,s_2)\rangle= \sqrt{2}G_FN_e.
\end{equation}

\section{Derivation of $V_Z^n$}\label{VZn} 
We now set out to find $V_Z^n$ due to $\nu_\alpha n$, $\alpha=e,\mu,\tau$ scattering. This reaction is mediated by the $Z^0$ boson, so we must use the effective Hamiltonian density Eq. (\ref{hz}). Now, the neutron consists of one $u$ and two $d$. The $u$ part of the Hamiltonian is 
\begin{align}\label{uq}
&\frac{G_F}{2\sqrt{2}}\bigl[\overline{u}(x)\gamma^\mu(1-\gamma_5 - \frac{8}{3}\sin^2\theta_W)u(x)\bigr]\bigl[\overline{\nu}_\alpha(x)\gamma_\mu(1-\gamma_5)\nu_\alpha(x)\bigr],
\end{align}
while the $d$ part is 
\begin{align}\label{dq}
&-\frac{G_F}{2\sqrt{2}}\bigl[\overline{d}(x)\gamma^\mu(1-\gamma_5 - \frac{4}{3}\sin^2\theta_W)d(x)\bigr]\bigl[\overline{\nu}_\alpha(x)\gamma_\mu(1-\gamma_5)\nu_\alpha(x)\bigr].
\end{align}
These contributions are to be added in the ratio 1:2 to obtain the relevant Hamiltonian for $\nu_en$ scattering. In total, this gives
\begin{align}
\frac{G_F}{2\sqrt{2}}&\Big[\overline{\psi}_n\gamma^\mu\big[(1-\gamma_5 - \frac{8}{3}\sin^2\theta_W)- 2(1-\gamma_5 - \frac{4}{3}\sin^2\theta_W)\bigr]\psi_n\Big][\overline{\nu}_\alpha(x)\gamma_\mu(1-\gamma_5)\nu_\alpha(x)\Big] \notag \\
&= -\frac{G_F}{2\sqrt{2}}\bigl[\overline{\psi}_n\gamma^\mu(1-\gamma_5)\psi_n\bigr]\bigl[\overline{\nu}_\alpha(x)\gamma_\mu(1-\gamma_5)\nu_\alpha(x)\bigr].
\end{align}
We are left with the effective Hamiltonian 
\begin{align}\label{NCham}
{\cal{H}}_Z(x) &= -\frac{G_F}{2\sqrt{2}} \int f(E_n,T)\times\frac{1}{2}\sum_s \bigl[\;\overline{\psi}_n(x) \gamma^\mu (1-\gamma_5)\psi_n(x)\bigr]\bigl[\overline{\nu}_\alpha(x)\gamma_\mu(1-\gamma_5)\nu_\alpha(x)\bigr] \mbox{d}{\bf p}_n.  
\end{align}
Here, we have introduced the statistical Fermi distribution for neutrons $f(E_n,T)$ and summation over the neutron spins due to the assumption of unpolarized medium, just as for the electrons. We see that Eq. (\ref{NCham}) is of the same form as Eq. (\ref{hwe}) if we use the earlier mentioned Fierz identity, such that the rest of the analysis is equivalent to the derivation of $V_W^e$. Following the same procedure as above, we find

\begin{align}\label{vzn11}
V_Z^n &= -\langle \nu_\alpha(p_1,s_1)n(p_2,s_2)| \frac{G_FN_n}{2\sqrt{2}}\times \int \overline{\nu}_\alpha(x)\gamma_0(1-\gamma_5)\nu_\alpha(x) \mbox{d}{\bf x}  |\nu_\alpha(p_1,s_1)n(p_2,s_2) \rangle \notag \\
&= -\langle \nu_\alpha(p_1,s_1)n(p_2,s_2)|\frac{G_FN_n}{2\sqrt{2}}\times \frac{1}{2VE_{\nu_\alpha}} \int \mbox{Tr}\{ (\pslash_{\nu_\alpha} + m_{\nu_\alpha})\gamma^0(1-\gamma^5) \} \notag\\
&\hspace{0.2in} \times \overbrace{a_{s_1}^\dagger({\bf p}_1) a_{s_1}({\bf p}_1)}^{\mbox{\scriptsize Num}_{s_1}({\bf p}_1)}    \mbox{d}{\bf x} |\nu_\alpha(p_1,s_1)n(p_2,s_2) \rangle\notag \\
&= -\langle \nu_\alpha(p_1,s_1)n(p_2,s_2)|\frac{G_FN_n}{2\sqrt{2}}\times \frac{1}{2VE_{\nu_\alpha}} \int 4E_{\nu_\alpha} \mbox{d}{\bf x} |\nu_\alpha(p_1,s_1)n(p_2,s_2)\rangle.
\end{align}

With normalized state vectors $|\nu_\alpha(p_1,s_1)n(p_2,s_2)\rangle$, Eq. (\ref{vzn11}) leads to
\begin{align}\label{finalvnz}
V_Z^n &= -\frac{G_FN_n}{2\sqrt{2}}\times \frac{2}{V} \int \mbox{d}{\bf x} \langle \nu_\alpha(p_1,s_1)n(p_2,s_2)||\nu_\alpha(p_1,s_1)n(p_2,s_2)\rangle= -\frac{G_FN_n}{\sqrt{2}}.
\end{align}

\section{Derivation of $V^e_Z$}\label{VZe}
Now, the expression for $V^e_Z$ is found by including the relevant terms from Eq. (\ref{hz}), which for NC $\nu_ee$ scattering reads
\begin{align}
-\frac{G_F}{2\sqrt{2}}\bigl[\overline{\nu}_e(x)\gamma^\mu(1-\gamma_5)\nu_e(x)\bigr]\bigl[\overline{e}(x)\gamma^\mu(1-\gamma_5-4\sin^2\theta_W)e(x)\bigr].
\end{align} 
Taking into account the Fermi distribution of electrons and the averaging of spins, the expression for $V^e_Z(x)$ takes the form 
\begin{align}
{\cal{H}}_Z(x) &= -\frac{G_F}{2\sqrt{2}} \int f(E_e,T)\times\frac{1}{2}\sum_s \bigl[ \overline{\nu}_e(x) \gamma^\mu (1-\gamma_5)\nu_e(x)\bigr]\bigl[\overline{e}(x)\gamma_\mu(1-\gamma_5-4\sin^2\theta_W)e(x)\bigr] \mbox{d}{\bf p}_e.  
\end{align}
As before, the rest of the analysis is equivalent to the derivation of $V_W^e$. In analogy to Eq. (\ref{vcc5}), it is found that 

\begin{align}\label{vzn1}
V_Z^e &= -\langle \nu_e(p_1,s_1)e(p_2,s_2)| \frac{G_FN_e}{2\sqrt{2}}\times \int (1-4\sin^2\theta_W)\overline{\nu}_\alpha(x)\gamma_0(1-\gamma_5)\nu_\alpha(x) \mbox{d}{\bf x}  |\nu_e(p_1,s_1)e(p_2,s_2) \rangle \notag \\
&= -\langle \nu_\alpha(p_1,s_1)e(p_2,s_2)|\frac{G_FN_e}{2\sqrt{2}}\times \frac{1}{2VE_{\nu_e}} \int (1-4\sin^2\theta_W)\mbox{Tr}\{ (\pslash_{\nu_e} + m_{\nu_e})\gamma^0(1-\gamma^5) \} \notag\\
&\hspace{0.2in} \times \overbrace{a_{s_1}^\dagger({\bf p}_1) a_{s_1}({\bf p}_1)}^{\mbox{\scriptsize Num}_{s_1}({\bf p}_1)}    \mbox{d}{\bf x} |\nu_e(p_1,s_1)e(p_2,s_2) \rangle\notag \\
&= -\langle \nu_e(p_1,s_1)e(p_2,s_2)|\frac{G_FN_e}{2\sqrt{2}}\times \frac{1}{2VE_{\nu_e}} \int (1-4\sin^2\theta_W)4E_{\nu_e} \mbox{d}{\bf x} |\nu_e(p_1,s_1)e(p_2,s_2)\rangle.
\end{align}
With normalized vectors $|\nu_e(p_1,s_1)n(p_2,s_2)\rangle$, Eq. (\ref{vzn1}) gives
\begin{align}\label{finalvez}
V_Z^e &= -\frac{G_FN_e}{2\sqrt{2}}\times \frac{2}{V}(1-4\sin^2\theta_W)\int \mbox{d}{\bf x} \langle \nu_e(p_1,s_1)e(p_2,s_2)||\nu_e(p_1,s_1)e(p_2,s_2)\rangle \notag\\
&= - \frac{G_F(1-4\sin^2\theta_W)N_e}{\sqrt{2}}.
\end{align}

\section{Derivation of $V^p_Z$}\label{VZp}
In order to derive $V^p_Z$ we attack the problem in the same way as for $V^n_Z$. A proton consists of two $u$ and one $d$ quarks, so first we find consider the individual contributions from each quark. From Eqs. (\ref{uq}) and (\ref{dq}) the total contribution to the proton is seen to be
\begin{align}
&\frac{G_F}{2\sqrt{2}}\Big[\overline{\psi}_p\gamma^\mu\big[2(1-\gamma_5 - \frac{8}{3}\sin^2\theta_W)-(1-\gamma_5 - \frac{4}{3}\sin^2\theta_W)\big]\psi_p\Big]\Bigl[\overline{\nu}_\alpha(x)\gamma_\mu(1-\gamma_5)\nu_\alpha(x)\Bigr] \notag\\
&\hspace{0.1in}= \frac{G_F}{2\sqrt{2}}\bigl[\overline{\psi}_p\gamma^\mu(1-\gamma_5-4\sin^2\theta_W)\psi_p\bigr]\bigl[\overline{\nu}_\alpha(x)\gamma_\mu(1-\gamma_5)\nu_\alpha(x)\bigr],
\end{align}
when added in the ratio 2:1, and including the extra factor of 2 due to the double quark contribution from Eq. (\ref{hz}). This gives the effective Hamiltonian for $\nu_\alpha p$, $\alpha=e,\mu,\tau$ scattering, namely
\begin{align}
{\cal{H}}_Z^p(x) &= \frac{G_F}{2\sqrt{2}} \int f(E_p,T)\times\frac{1}{2}\sum_s \bigl[ \overline{\psi}_p(x) \gamma^\mu (1-\gamma_5-4\sin^2\theta_W)\psi_p(x)\bigr]\bigl[\overline{\nu}_\alpha(x)\gamma_\mu(1-\gamma_5)\nu_\alpha(x)\bigr] \mbox{d}{\bf p}_n.  
\end{align}
The further analysis is then just as for $V_W^e$, and leads to
\begin{align}\label{vzp1}
V_Z^p &= -\langle \nu_\alpha(p_1,s_1)p(p_2,s_2)| \frac{G_FN_p}{2\sqrt{2}}\times \int (1-4\sin^2\theta_W)\overline{\nu}_\alpha(x)\gamma_0(1-\gamma_5)\nu_\alpha(x) \mbox{d}{\bf x}  |\nu_\alpha(p_1,s_1)p(p_2,s_2) \rangle \notag \\
&= \langle \nu_\alpha(p_1,s_1)p(p_2,s_2)|\frac{G_FN_p}{2\sqrt{2}}\times \frac{1}{2VE_{\nu_\alpha}} \int (1-4\sin^2\theta_W) \mbox{Tr}\{ (\pslash_{\nu_\alpha} + m_{\nu_\alpha})\gamma^0(1-\gamma^5) \} \notag\\
&\hspace{0.2in} \times \overbrace{a_{s_1}^\dagger({\bf p}_1) a_{s_1}({\bf p}_1)}^{\mbox{\scriptsize Num}_{s_1}({\bf p}_1)}    \mbox{d}{\bf x} |\nu_\alpha(p_1,s_1)p(p_2,s_2) \rangle\notag \\
&= \langle \nu_\alpha(p_1,s_1)p(p_2,s_2)|\frac{G_FN_p}{2\sqrt{2}}\times \frac{1}{2VE_{\nu_\alpha}} \int (1-4\sin^2\theta_W)4E_{\nu_\alpha} \mbox{d}{\bf x} |\nu_\alpha(p_1,s_1)p(p_2,s_2)\rangle.
\end{align}

As before, we require the state vectors $|\nu_\alpha(p_1,s_1)p(p_2,s_2)\rangle$ to be normalized. Eq. (\ref{vzp1}) then gives
\begin{align}\label{finalvpz}
V_Z^p &= \frac{G_FN_p}{2\sqrt{2}}\times \frac{2}{V}(1-4\sin^2\theta_W)\int \mbox{d}{\bf x} \langle \nu_\alpha(p_1,s_1)p(p_2,s_2)||\nu_\alpha(p_1,s_1)p(p_2,s_2)\rangle \notag\\
&= \frac{G_FN_p(1-4\sin^2\theta_W)}{\sqrt{2}}.
\end{align}

\section{Justification of treating neutrons and protons as point-particles}
At first sight, it might seem startling that the neutron and proton are treated as point particles with respect to the neutrino scattering, instead of taking the well-known nucleon form factors into account. But it turns out that these are actually irrelevant as far as the matter potentials are concerned. Notice how all NC potentials are directly proportional to the corresponding vector coupling constant $g_V$ of the scattering particle, i.e. $N_e \propto g_V^e = 2\sin^2 \theta_W - 1/2$, $N_n \propto g_V^n = -1/2$, $N_p \propto g_V^p = 1/2 - 2\sin^2 \theta_W$. The nucleon vector coupling constants are obtained by quite simply adding the quark equivalents for the $u$ and $d$ in the correct ratio; 2:1 for $p$ and 1:2 for $n$. \\

An attempt to apply the same logic to the axial coupling constant $g_A$ would not be successful. The axial current anomaly (see {\it e.g.} Ref. \cite{peskin}) modifies the value of this quantity, thus yielding for instance $g_V^p \simeq 1.27/2$ instead of 1/2 (this particular number has been determined by experiments, see {\it e.g.} Ref. \cite{proton1}). One compensates for this fact by operating with form-factors in scattering reactions on nucleons. 
For neutrino forward scattering (essentially the MSW-effect), one needs to include form factors that are non-vanishing in the limit of zero momentum transfer. Because of the mentioned axial current anomaly, the form factor for the axial current will give results different from a combination of quark currents in the respective ratios for neutrons and protons. But the potentials were generally derived to be independent of $g_A$, regardless of scattering component. As a consequence, we need not worry about the axial form-factors. Left with only the vector part, the nucleon can simply be treated as the superposition of three quarks in this scenario. Of course, such an uncomplicated picture is certainly the exception rather than the rule with respect to nucleon scattering.

\section{How to treat a varying density}
Eqs. (\ref{finalvnz}), (\ref{finalvez}), (\ref{finalvpz}) are equally valid even if the particle density $N_i$ fluctuates in space. One would compensate for such behavior by evaluating the matter potential at each separate point in an artifical lattice-space, or alternatively approximating the particle density with a contineous function of the space-coordinate $\mathbf{x}$. The two cases constant vs. non-constant particle density differ crucially in terms of how they affect the probability for a neutrino oscillation to have occured (see {\it e.g.} Ref. \cite{msw2} for an introduction to the topic of neutrino oscillations). Dependent on which of the two scenarios that is applicable, the oscillation probability takes on a certain characteristic spatial-dependence. This is also dependent on the nature of the particle density fluctuations, i.e. adiabiatic, unit-step profile, et.c. However, oscillations are not the topic of this paper, and we shall settle with the derivation of Eqs. (\ref{finalvnz}), (\ref{finalvez}), (\ref{finalvpz}) since these are heuristically valid even in the case of a varying particle density.

\section{Potential sign for antineutrinos}
The matter potentials derived so far has been for neutrinos. One cannot generalize these results to be valid for antineutrinos, and here is why. Consider the induced matter potential for $\nu_e$ due to NC scattering on electrons. This potential is given as (see Sec. \ref{VZe})
\begin{equation}\label{neuex}
V_Z^e = \langle \nu_e(p_1,s_1)e(p_2,s_2)|H_Z^e|\nu_e(p_1,s_1)e(p_2,s_2) \rangle,
\end{equation}
while the induced matter potential for $\overline{\nu}_e$ for the same type of reaction reads
\begin{equation}\label{antineuex}
\overline{V}_Z^e = \langle \overline{\nu}_e(p_1,s_1)e(p_2,s_2)|H_Z^e|\overline{\nu}_e(p_1,s_1)e(p_2,s_2) \rangle.
\end{equation}
In the general case, the operator part of Eq. (\ref{antineuex}) takes the form
\begin{equation}\label{mH}
\langle 0| b_{\nu}({\bf p}) \Big[ \sum_{\bf k,k'} b_{\nu}({\bf k}) b_{\nu}^\dag({\bf k'}) \Big] b_{\nu}^\dag({\bf p'}) |0\rangle,
\end{equation}
where the sum over $b_{\nu}({\bf k}) b_{\nu}^\dag({\bf k'})$ comes from the second quantized fields $\overline{\nu}_e(x)\nu_e(x)$. Using the commutation relation $[b_{\nu}(\mathbf{k}), b_{\nu}^\dag(\mathbf{k'})] = \delta(\mathbf{k'} - \mathbf{k})$, Eq. (\ref{mH}) becomes
\begin{equation}\label{mH1}
-\langle 0| b_{\nu}({\bf p}) \Big[ \sum_{\bf k,k'} b_{\nu}^\dag({\bf k'}) b_{\nu}({\bf k}) \Big] b_{\nu}^\dag({\bf p'}) |0\rangle.
\end{equation}
The only non-vanishing contribution from this term is obtained by taking $\mathbf{k} = \mathbf{p'}, \mathbf{k'} = \mathbf{p}$. In our case of elastic scattering, we impose the limit $\mathbf{p} \to \mathbf{p'}$. A similar argument for Eq. (\ref{neuex}) produces the operator sequence
\begin{equation}\label{mH3}
\langle 0| a_{\nu}({\bf p}) \Big[ \sum_{\bf k,k'} a_{\nu}^\dag({\bf k'}) a_{\nu}({\bf k}) \Big] a_{\nu}^\dag({\bf p'}) |0\rangle.
\end{equation}
Here, $a_{\nu}({\bf p})$ and $a_{\nu}^\dag({\bf p})$ ($b_{\nu}({\bf p})$ and $b_{\nu}^\dag({\bf p})$) are interpreted as annihilation and creation operators for neutrinos (antineutrinos). The matter potentials differ in sign due to the relative minus sign between Eqs. (\ref{mH1}) and (\ref{mH3}). Thus, $\overline{V}_Z^e = - V_Z^e$. \\

\section{Summary}
The relevant matter potentials for neutrinos propagating through Earth have been derived using a method that applies in the same way to both leptons and nucleons. We have explicitely shown why the matter potential sign is reversed in the case of antineutrinos. Specifically, it is worth to note that the effective potential in an electrically neutral medium with respect to neutrino oscillations is given by $H_W^e$ in Eq. (\ref{eq:hmsw}). From Eqs. (\ref{finalvez}) and (\ref{finalvpz}) it is easily seen that
\begin{equation}
V^e_Z + V^p_Z \Big|_{N_e = N_p} = 0.
\end{equation}
Thus, there is no effective matter potential felt by neutrinos due to scattering on protons and electrons mediated by $Z^0$ in an electrically neutral medium. Also, since $H_Z^e$ in Eq. (\ref{eq:hmsw}) is diagonal, it gives an overall phase shift to all neutrino flavors, which is of no significance in the oscillation scenario. Our results are summarized in Tab. \ref{tab:matterpotentials}.
\begin{center}
\begin{table}[h!]
\centering{
	\caption{Neutrino matter potentials induced by Earth.}
	\label{tab:matterpotentials}
	\vspace{0.15in}
	\begin{tabular}{cc}
		 \hline
		 \hline
		 {\bf Type of reaction}	\hspace{0.15in}	& \bf Matter potential 				\\
	  	 \hline
		 $V_Z^n$		\hspace{0.15in} & $\mp G_FN_n/\sqrt{2}$	 			\\
		 $V_Z^p$		\hspace{0.15in} & $\pm G_F(1-4\sin^2\theta_W)N_p/\sqrt{2}$  	\\
		 $V_Z^e$		\hspace{0.15in} & $\mp G_F(1-4\sin^2\theta_W)N_e/\sqrt{2}$       \\		
		 $V_W^e$		\hspace{0.15in} & $\pm \sqrt{2}G_FN_e$	                 \\	
		 \hline
		 \hline
	\end{tabular}}
\end{table}
\end{center}
Generalization to matter potentials such as $V_Z^\mu$ are obtained simply by substituting $N_e \to N_\mu$. The upper sign refer to neutrinos, while the lower gives the matter potential for antineutrinos.

\end{document}